\documentclass[twocolumn]{article}

\usepackage{soul}
\usepackage[utf8]{inputenc}
\usepackage{listings}
\usepackage{lipsum}
\usepackage{caption} 
\usepackage{xcolor}
\usepackage{graphicx}
\usepackage{url}
\usepackage{hyperref}
\usepackage{authblk}
\usepackage{orcidlink}

\definecolor{codegreen}{rgb}{0,0.6,0}
\definecolor{codegray}{rgb}{0.5,0.5,0.5}
\definecolor{codepurple}{rgb}{0.58,0,0.82}
\definecolor{backcolour}{rgb}{0.95,0.95,0.95}

\title{bio2Byte Tools deployment as a Python package and Galaxy tool to predict protein biophysical properties}

\author[1, 2, \dag]{Jose Gavalda-Garcia\orcidlink{0000-0001-6431-3442}}
\author[1, 2, \dag]{Adrián Díaz\orcidlink{0000-0003-0165-1318}}
\author[1, 2, $\ast$]{Wim Vranken\orcidlink{0000-0001-7470-4324}}

\affil[1]{Interuniversity Institute of Bioinformatics in Brussels, ULB-VUB, Brussels, Belgium}
\affil[2]{Structural Biology Brussels, Vrije Universiteit Brussel, Brussels, Belgium}

\affil[$\dag$]{Authors contributed equally to this work. }
\affil[$\ast$]{Corresponding author. \href{email:wim.vranken@vub.be}{wim.vranken@vub.be}}

\begin{document}
\twocolumn[
  \begin{@twocolumnfalse}
    \maketitle
    \abstract{
\textbf{Summary:} We introduce a unified Python package for the prediction of protein biophysical properties, streamlining previous tools developed by the Bio2Byte research group. This suite facilitates comprehensive assessments of protein characteristics, incorporating predictors for backbone and sidechain dynamics, local secondary structure propensities, early folding, long disorder, beta-sheet aggregation and FUS-like phase separation. Our package significantly eases the integration and execution of these tools, enhancing accessibility for both computational and experimental researchers.\\
\textbf{Availability and Implementation:} The suite is available on the Python Package Index (PyPI): \url{https://pypi.org/project/b2bTools/} and Bioconda: \url{https://bioconda.github.io/recipes/b2btools/README.html} for Linux and macOS systems, with Docker images hosted on Biocontainers: \url{https://quay.io/repository/biocontainers/b2btools?tab=tags&tag=latest} and Docker Hub: \url{https://hub.docker.com/u/bio2byte}. Online deployments are available on Galaxy Europe: \url{https://usegalaxy.eu/root?tool_id=b2btools_single_sequence} and our online server: \url{https://bio2byte.be/b2btools/}. The source code can be found at \url{https://bitbucket.org/bio2byte/b2btools_releases}.\\
\textbf{Contact:} \href{wim.vranken@vub.be}{wim.vranken@vub.be}
}
    \vspace{1cm}
  \end{@twocolumnfalse}
]

\section{Introduction}

Proteins are complex molecules whose motions often play a fundamental role in their function \cite{fenwick_integrated_2014, campbell_role_2016}. The experimental study of the dynamics of proteins offers insights on properties such as order and flexibility, folding mechanics, conformational changes and secondary structure populations \cite{berjanskii_nmr_2006, eaton_modern_2021}. Such experiments, often using Nuclear Magnetic Resonance (NMR), provide valuable protein dynamics information but are expensive and time consuming with no high throughput possible. To obtain proteome-scale estimations of such characteristics, predictors are essential, for example to examine trends between how expressable protein fragments are and their dynamics \cite{boone_massively_2021}, or to inform protein early folding characteristics in relation to experiments  \cite{smets_evolutionary_2022}.

We developed an assortment of predictors of protein properties that work from single amino acid sequences, encompassing estimations of backbone dynamics \cite{cilia_protein_2013, cilia_dynamine_2014}, early folding \cite{raimondi_exploring_2017} and long disorder \cite{orlando_prediction_2022} among others. The code for these tools was often developed separately, illustrating the natural progression of scientific software, with each tool originally employing its own contemporary dependencies and programming language versions. Though the code was openly available and each tool could be individually compiled with the precise set of dependencies' versions described in their repositories, this effectively requires a separate environment per tool and, potentially, significant effort to make each tool work. This contrasts with the complimentary nature of these tools, which together offer a more holistic vision of a protein's (dynamic) nature.

To facilitate the use of our tools, we created and published a web server where users can submit protein sequences and obtain the output of our tools \cite{kagami_b2btools_2021}. These calculations are convenient from the user perspective, but are limited in throughput due to server constraints, as well as being susceptible to interruptions due to restricted computational resources.

\begin{figure*}[!t]%
	\includegraphics[width=\linewidth]{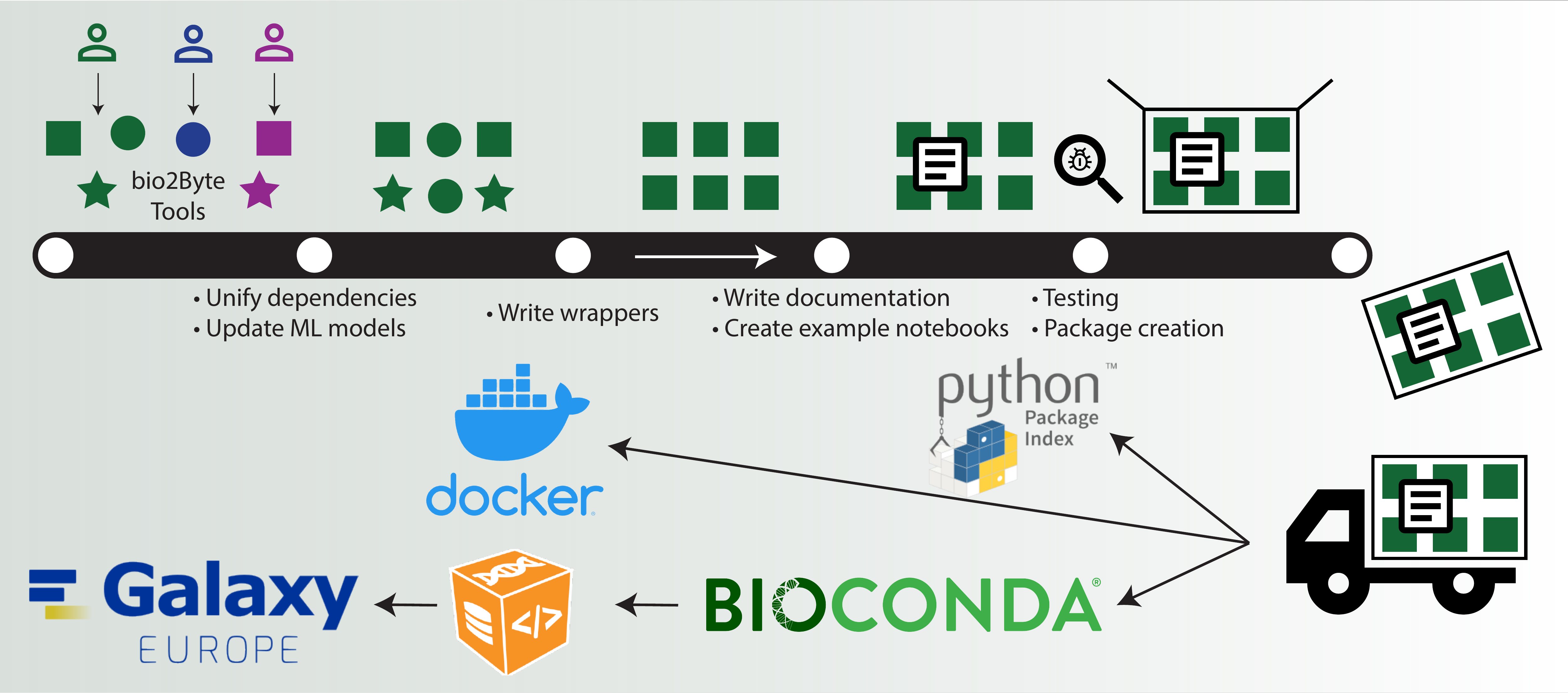}  
	\centering
	\caption{\textbf{Overview of the update and deployment process applied to bio2Byte Tools.} Even within a single research group, the development of tools over time and across various developers carries variation in dependencies and coding conventions, illustrated in the heterogeneous shapes and colours for the original softwares (top left). These tools have to go through a (figurative) production line to become a harmonised software package, during which the different tools acquire compatible shapes and colours. Once the tools fit together, a user manual, an extensive README.md and example notebooks can be included. Finally, quality is assured in the form of thorough execution and outputs tests and the package with all the harmonised tools can be created. This package leaves the production line and is deployed to different distribution channels and archived for ease of use and reproducibility.}
	\label{overview_figure}
\end{figure*}

Here we present a unified deployment of our tools as a single Python package that can be employed on any Linux or macOS machine. The deployment protocol that we have defined publishes the package on the Python Package Index (PyPI) \cite{pypi} and Bioconda, the biomedical research channel of Conda package manager (only for Linux 64-bit and AArch64 or macOS x86 64)\cite{gruning_bioconda_2018}. In addition, it archives on Biocontainers \cite{da_veiga_leprevost_biocontainers_2017} to ensure reproducibility and publishes it on Galaxy \cite{afgan_galaxy_2018} Europe to enable its incorporation into pipelines and facilitating non-programmatic usage (process overview in figure \ref{overview_figure}). 

\section{Tools harmonisation and packaging}
The tools included in this version of bio2Byte Tools were primarily sequence-based predictors, namely DynaMine \cite{cilia_protein_2013}, DisoMine \cite{orlando_prediction_2022}, EfoldMine \cite{raimondi_exploring_2017}, AgMata \cite{orlando_accurate_2020} and PSPer \cite{orlando_computational_2019}, which can estimate an array of metrics from a simple FASTA file, but also includes ShiftCrypt \cite{orlando_shiftcrypt_2020} for residue-level interpretation of NMR Chemical Shift data for proteins (Supplementary table \ref{tab:tool_explanation}). Here, we focus on the description and integration of the sequence-based predictors, which facilitates their deployment in a single environment. The steps described in this work result in further FAIRification of our tools \cite{wilkinson_fair_2016}, now robustly available for the scientific community.

\subsection{Code and models updating}
The fundamental first step to transform a set of tools into a suite was to unify all the dependencies. The code of all tools was updated to adopt syntax compatible with all Python versions between Python 3.7 and 3.12 and the deprecated calls to objects in external dependencies were substituted for current code with equivalent behaviour. Notably, the machine learning model files of these tools had to be updated to match the format of the updated dependencies, an action that is not recommended by the employed machine learning libraries (Scikit-learn \cite{scikit-learn} and PyTorch \cite{paszke_pytorch_2019}) and therefore, no standard model conversion procedure exists. This step required extensive labour to understand the old and updated model description file formats, to then update the trained model files. Given the risk to change the output of the tools, thorough tests were implemented in a diverse set of proteins. This ensured that the difference in outputs between legacy and unified tools was no larger than floating point errors.

\subsection{Programmatic usage and execution}

To facilitate the execution of the predictors inside the package, their inner classes and methods were abstracted in a public wrapper class, named ``\textit{SingleSeq}'' and illustrated in code snippet \ref{single_seq_snippet}, which is the entry point for the users and acts as the interface between them and our predictors. This wrapper allows an easy parse of sequences in FASTA format, execution of the desired tools, storage for the outputs as a Python object dictionary, and export outputs and metadata as a JSON, CSV or TSV file (exemplified in code snippet \ref{single_seq_snippet} for Python usage and code snippet \ref{single_seq_snippet_bash} for bash usage). An equivalent wrapper is also available to execute the tools on a multiple sequence alignment (MSA), named ``\textit{MultipleSeq}'' and described in README.md, which maintains the aligned position of the amino acids and returns the bio2Byte Tools outputs indexed according to this alignment.

\begin{lstlisting}[caption={Python execution of bio2Byte Tools. This code will store for each individual sequence contained in \textit{example.fasta} a \textit{.json} output file with the predictions of the selected tools and their dependencies.}, label={single_seq_snippet}, float] 
	import json
	from b2bTools import SingleSeq
	
	input_fasta = "/path/to/example.fasta"
	
	single_seq = SingleSeq(input_fasta)
	single_seq.predict(tools=["disomine"])
	preds = single_seq.get_all_predictions()
	
	for seq_id, pred_vals in preds["proteins"]:
	with open(f"{seq_id}.json", "w") as fp:
	json.dump(pred_vals, fp)
\end{lstlisting}

\begin{lstlisting}[language=bash, caption={Bash execution of bio2Byte Tools. This code will store the output of all sequences in the \textit{.fasta} input in \textit{.json} and \textit{.csv} files and will also store the metadata in a different file.}, label={single_seq_snippet_bash}, float] 
	b2bTools \
	--input_file /input/example.fasta \
	--output_json_file /output/example.json \
	--output_tabular_file /output/example.csv \
	--metadata_file /output/example.meta.csv
\end{lstlisting}

To optimise resources, when a user selects the set of tools to run, their interdependencies are assessed (supplementary table \ref{tab:tool_dependencies}) and all required tools will be executed only once in the necessary order. For example, if DisoMine and DynaMine are selected for execution, DynaMine will run first, then EfoldMine (a requirement for DisoMine) and finally, DisoMine. In contrast, the legacy code would execute multiple instances of DynaMine, for each tool that depended on it. 


\subsection{Testing and publishing}

Our tools were then contained in a Python package. With every new version, this package is subject to exhaustive unit tests using the library \lstinline|pytest| \cite{pytestx.y} to ensure codebase integrity and predictions consistency, \textit{i.e.} we assert that the code modification does not alter the predictor's outputs. 
We have automated the building and testing process by including a CD/CI pipeline in the package repository on Bitbucket. We run in parallel all the unit tests for every compatible Python version. Then the pipeline builds the Python package and tests its command-line execution with a simple FASTA input file (supplementary figure \ref{fig:pytest}).

If all tests are successful, a new version is manually published on PyPI and Bioconda, which additionally tests that all modules and sub-modules can be correctly imported and called. Once these tests are passed, the new version of bio2Byte Tools get incorporated in Bioconda's index. As soon as this new version is included in Bioconda's index, a new Docker image is automatically built and published on the Biocontainers containers registry. 
This image on Biocontainers is created only for Python 3.10, so we also create a set of Docker images for each new version in three different flavours for each Python from 3.7 to 3.12 (supplementary table \ref{tab:docker_versions}). Each of these containers are subject to an additional iteration of the previously described unit tests upon creation. This ensures system-independent execution and reproducibility of results for every version of our software.

\section{Deployment in online pipelines}
With the software available online as a Python package and as a Docker image, we uploaded it as a tool on Galaxy Europe. This enables its usage in this platform as a standalone tool or as part of a workflow with other tools. The creation of the Galaxy package required the definition of the tool's user interface, which then defines all necessary settings to run our tools. This user interface allows code-free execution and, if selected, plotting of its outputs.

\section{Future perspectives}

The docker image of bio2Byte Tools can now be deployed on our own server \cite{kagami_b2btools_2021}, and will soon substitute the original custom and more convoluted approach with multiple environments. 

Furthermore, one of the most anticipated features under development for this software suite is the creation of a protocol for Nextflow \cite{di_tommaso_nextflow_2017} so that our tools can be easily incorporated and shared as part of reproducible pipelines. Nextflow also allows for easy scaling of the workload among all available processing units, effectively accelerating the calculation of large data sets. 

Additionally, though this suite is able to use evolutionary information from an MSA to estimate the biophysical constraints of residue positions in a protein family, it does not provide additional interpretation of such results. We are currently working on an extra layer of interpretation for such information, for example to determine how well a protein fits within those biophysical constraints. 

We will continue to update the suite as new Python versions are released, aiming to maintain support for older versions whenever possible. Finally, we hope the improved usability and reproducibility of our tools will encourage its further incorporation in scientific resources.

\section{Additional resources for bio2Byte Tools}

An extensive collection of example notebooks can be found on \url{https://github.com/Bio2Byte/public_notebooks/}, which will reflect future updates and new functionalities of our suite.

\section{Conflicts of interest}
The authors declare no conflict of interest.

\section{Author contributions}
J.G.-G. and A.D. developed, implemented and validated the software suite and CI/CD pipelines.
W.V. provided supervision and implemented the initial integrated version of the prediction tools.
All authors contributed to the writing of the manuscript.

\section{Funding}
This work has been supported by the European Union’s Horizon 2020 research and innovation program under the Marie Skłodowska-Curie grant agreement [813239 to  J.G.-G. and A.D.]; Research Foundation Flanders (FWO) International Research Infrastructure [I000323N to W.V. and G.0328.16N to A.D.]; COST Action ML4NGP, CA21160, supported by COST (European Cooperation in Science and Technology).

\appendix

\onecolumn
\section*{Supplementary information}

\setcounter{table}{0}
\setcounter{figure}{0}

\captionsetup[table]{name=Supplementary Table}
\captionsetup[figure]{name=Supplementary Figure}

\begin{table}[ht]
	\centering
	\caption{\textbf{Tools description.} Simple description of every tool contained in bio2Byte tools.}
	\label{tab:tool_explanation}
	\begin{tabular}{|c|p{10cm}|}
		\hline
		\textbf{Predictor} & \textbf{Usage} \\ \hline
		DynaMine & Fast predictor of protein backbone dynamics using only sequence information as input. The current version also predicts side-chain dynamics and secondary structure predictors using the same principle. \\ \hline
		DisoMine & Predicts protein disorder with recurrent neural networks not directly from the amino acid sequence, but instead from more generic predictions of key biophysical properties, here protein dynamics, secondary structure and early folding. \\ \hline
		EfoldMine & Predicts from the primary amino acid sequence of a protein, which amino acids are likely involved in early folding events. \\ \hline
		AgMata & Single-sequence based predictor of protein regions that are likely to cause beta-aggregation. \\ \hline
		PSPer & PSP (Phase Separating Protein) predicts whether a protein is likely to phase-separate with a particular mechanism involving RNA interacts (FUS-like proteins). It will highlight protein regions that are involved mechanistically, and provide an overall score. \\ \hline
		ShiftCrypt & Auto-encoding NMR chemical shifts from their native vector space to a residue-level biophysical index. \\ \hline
	\end{tabular}
\end{table}

\begin{table}[ht]
	\centering
	\caption{\textbf{Overview of internal tool dependencies.} The predictors contained in this software suite often require the execution of other predictors in our suite. The optimisation process is explained in the main text and the list of dependencies is described in this table. }
	\label{tab:tool_dependencies}
	\begin{tabular}{|l|l|}
		\hline
		\textbf{Tool} & \textbf{Dependencies} \\ \hline
		DynaMine & None \\ \hline
		EFoldMine & DynaMine \\ \hline
		DisoMine & DynaMine, EFoldMine \\ \hline
		AGmata & DynaMine, EFoldMine \\ \hline
		PSPer & DynaMine, EFoldMine, DisoMine\\ \hline
		ShiftCrypt & None \\ \hline
	\end{tabular}
\end{table}

\begin{table}[ht]
	\centering
	\caption{\textbf{Python Versions and Distribution Sources for all the manually generated Docker images.} For each new release of bio2Byte Tools, the following set of Docker images are generated and published on \url{https://hub.docker.com/r/bio2byte/b2btools} for additional control over the version of Python and the provenance of the dependencies.}
	\label{tab:docker_versions}
	\begin{tabular}{|p{1.5cm}|l|p{7cm}|}
		\hline
		\textbf{Python Version} & \textbf{Source} & \textbf{Description} \\ \hline
		3.7 & PyPI & Generated from PyPI distributions using Python 3.7. \href{https://hub.docker.com/layers/bio2byte/b2btools/3.0.7b3-pypi_py3.7-linux64/images/sha256-5327a11cd680114080ab3a7e7cea1917e45536a3816c4c3baddb2d981b4cfb55?context=explore}{(Link to image)} \\ \hline
		3.7 & Conda & Generated from Conda distributions using Python 3.7. \href{https://hub.docker.com/layers/bio2byte/b2btools/3.0.7b3-conda_py3.7-linux64/images/sha256-c8ffc40d4eda5449721bbb8aa74a2fcc7a873843c84a4ee585354a53177b3e25?context=explore}{(Link to image)} \\ \hline
		3.7 & System & Generated from System packages using Python 3.7. \href{https://hub.docker.com/layers/bio2byte/b2btools/3.0.7b3-pkg_py3.7-linux64/images/sha256-063cd6df621017c0194a52c7b8116db800dc525f2bf380e7e2f6966b01f4daef?context=explore}{(Link to image)} \\ \hline
		3.8 & PyPI & Generated from PyPI distributions using Python 3.8. \href{https://hub.docker.com/layers/bio2byte/b2btools/3.0.7b3-pypi_py3.8-linux64/images/sha256-13537a7fc7289fd3f85ddde7d992164bb6e7e28b9bd3c77656dffbd0db9d94d1?context=explore}{(Link to image)} \\ \hline
		3.8 & Conda & Generated from Conda distributions using Python 3.8. \href{https://hub.docker.com/layers/bio2byte/b2btools/3.0.7b3-conda_py3.8-linux64/images/sha256-905d3914d9774cfce596f6af027088a154ded1f95abae0d09e3c7e97fbf0d059?context=explore}{(Link to image)} \\ \hline
		3.8 & System & Generated from System packages using Python 3.8. \href{https://hub.docker.com/layers/bio2byte/b2btools/3.0.7b3-pkg_py3.8-linux64/images/sha256-6fb04d6c62f5023ad47471ee3cec7cd687dfdeaf89e9d8ba9bd75eade03f394e?context=explore}{(Link to image)} \\ \hline
		3.9 & PyPI & Generated from PyPI distributions using Python 3.9. \href{https://hub.docker.com/layers/bio2byte/b2btools/3.0.7b3-pypi_py3.9-linux64/images/sha256-5e0eeb86bf8189f34d22f7420e25f46811a416880e494dcc7d4601c2a2d66cc8?context=explore}{(Link to image)} \\ \hline
		3.9 & Conda & Generated from Conda distributions using Python 3.9. \href{https://hub.docker.com/layers/bio2byte/b2btools/3.0.7b3-conda_py3.9-linux64/images/sha256-aca1984afbb71132923dfca301b74d24a688a2584bfeddc24fa720bfc056bf42?context=explore}{(Link to image)} \\ \hline
		3.9 & System & Generated from System packages using Python 3.9. \href{https://hub.docker.com/layers/bio2byte/b2btools/3.0.7b3-pkg_py3.9-linux64/images/sha256-9f88a39914e50336e222f1dd1a8c1b6a9be05711f186533eaa7ea4856c066798?context=explore}{(Link to image)} \\ \hline
		3.10 & PyPI & Generated from PyPI distributions using Python 3.10. \href{https://hub.docker.com/layers/bio2byte/b2btools/3.0.7b3-pypi_py3.10-linux64/images/sha256-e0937fe94e181fb076353469c4e5098ae5a337121deadd4b3c13be23fdc37e6e?context=explore}{(Link to image)} \\ \hline
		3.10 & Conda & Generated from Conda distributions using Python 3.10. \href{https://hub.docker.com/layers/bio2byte/b2btools/3.0.7b3-conda_py3.10-linux64/images/sha256-ad02b57eae26760992053fc257bc67cc079f4c2d06be294cf7053defd2712bb5?context=explore}{(Link to image)} \\ \hline
		3.10 & System & Generated from System packages using Python 3.10. \href{https://hub.docker.com/layers/bio2byte/b2btools/3.0.7b3-pkg_py3.10-linux64/images/sha256-fdb6afe502361d61c2d38e0dd86ba301d6851ec206613a529a8752275a874e9a?context=explore}{(Link to image)} \\ \hline
		3.11 & PyPI & Generated from PyPI distributions using Python 3.11. \href{https://hub.docker.com/layers/bio2byte/b2btools/3.0.7b3-pypi_py3.11-linux64/images/sha256-8c7a49afcd780cf6466193ed0c20c7eb128256ecddeeaaebe85f25589ca275ae?context=explore}{(Link to image)} \\ \hline
		3.11 & Conda & Generated from Conda distributions using Python 3.11. \href{https://hub.docker.com/layers/bio2byte/b2btools/3.0.7b3-conda_py3.11-linux64/images/sha256-3e290ba287d254bc166a03fb66321370f24dc1408c060f6e92246061df46d519?context=explore}{(Link to image)} \\ \hline
		3.11 & System & Generated from System packages using Python 3.11. \href{https://hub.docker.com/layers/bio2byte/b2btools/3.0.7b3-pkg_py3.11-linux64/images/sha256-edb04f9a3fb1c6974db32abeeb274351bc15b9b04a503037919c2b5bbf3e9f9f?context=explore}{(Link to image)} \\ \hline
		3.12 & PyPI & Generated from PyPI distributions using Python 3.12. \href{https://hub.docker.com/layers/bio2byte/b2btools/3.0.7b3-pypi_py3.12-linux64/images/sha256-18060407097485c100b382da9651bff4a6e5b7e63d7ecb0669c11b32807d33b1?context=explore}{(Link to image)} \\ \hline
		3.12 & Conda & Generated from Conda distributions using Python 3.12. \href{https://hub.docker.com/layers/bio2byte/b2btools/3.0.7b3-conda_py3.12-linux64/images/sha256-f054968f96fe653515aa81481b67683473f924cbd479eecc114d91d2b43d3d39?context=explore}{(Link to image)} \\ \hline
		3.12 & System & Generated from System packages using Python 3.12. \href{https://hub.docker.com/layers/bio2byte/b2btools/3.0.7b3-pkg_py3.12-linux64/images/sha256-37c7dbc9efb4b054520d3ddd3768ce888cf87ed10116141f2bde48eb2e4a1e12?context=explore}{(Link to image)} \\ \hline
		
	\end{tabular}
\end{table}

\begin{figure*}[!t]%
	\includegraphics[width=\linewidth]{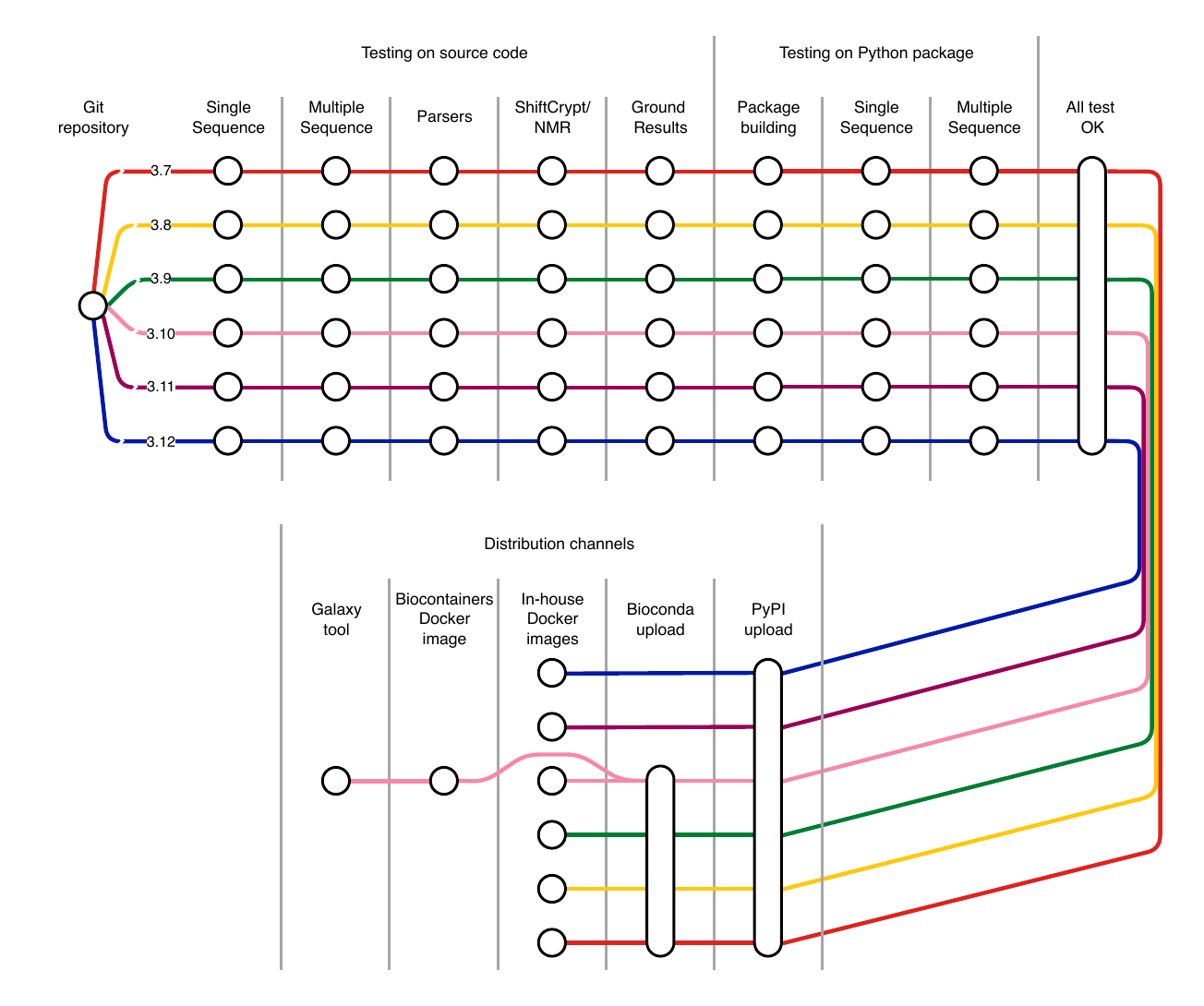}  
	\centering
	\caption{\textbf{Pytest process diagram.} The package testing process starts from a single Git repository, which then creates an environment per Python version from 3.7 to 3.12. For each environment, a series of tests on the source code are executed, namely the execution of single sequence, multiple sequence, parsers, and ShiftCrypt/NMR modules. Then the outputs get compared to ground results values and assessed for differences beyond float point errors (sum of absolute differences across all predicted metrics and residues, proteins and residues smaller than 0.001). A python package is then created and single sequence and multiple sequence modules are newly tested. If all tests across all Python versions are satisfactory, the python package will be uploaded to PyPI and to Bioconda. Then, an array of Docker images are created and uploaded to Docker Hub, following the settings in supplementary table \ref{tab:docker_versions}. Finally, Bioconda automatically deposits a docker container with the Python 3.10 version of the package in Biocontainers, which can then be deployed into Galaxy.}
	\label{fig:pytest}
\end{figure*}

\end{document}